\documentclass[letterpaper,10pt]{article} 
\usepackage{bm} 
\usepackage{tikz}
\usetikzlibrary{arrows.meta}
\usepackage{amsmath}

\newcommand{\pbm}{%
\begin{tikzpicture}
      \foreach \x in {2.7,2.9,3.1} {
        \draw[black, thick, fill=white] (\x,2.4) ellipse [x radius=0.2, y radius=0.4];
      }
      \node (xin) at (0.5,2) {};
      \node (yout) at (6.9,2) {};
      \draw[-{Latex[length=1.5mm]}, thick] (xin) -- (yout);
      \draw[thick, fill=white] (4.3+1.7*0.4,2) -- (4.3,2.4) -- (4.3,1.6) -- cycle;
    
      \node[anchor=east] at (xin) {\footnotesize $\vect{x}$};
      \node[anchor=west] at (yout) {\footnotesize $\vect{y}$};
      \node at (5, 2.7) {\footnotesize Optical fiber channel};

    \draw[-{Latex[length=1.5mm]}, thick] (0.5,0) -- (6.9,0);

    \draw[color=RoyalBlue, fill=RoyalBlue!10, rounded corners=3pt, thick] (1,-0.95) rectangle (6,1.25); \node[align=left] at (2.4, 1) {\footnotesize Parameterized SSFM};
    \draw[-{Latex[length=1.5mm]}, thick] (1.01,0) -- (1.2,0); \draw[-{Latex[length=1.5mm]}, thick] (3.75+0.775,0) -- (5.025,0); \draw[thick] (5.025+0.775,0) -- (5.99,0); \draw[-{Latex[length=1.5mm]}, thick] (1.2+0.775 ,0) -- (2.475,0);

    \foreach \x/\eq in {1.2/ $H({\cdot;\,\thetaDT^{(1)}})$, 2.475/$H({\cdot;\,\thetaDT^{(2)}})$, 3.75/$H({\cdot;\,\thetaDT^{(M-1)}})$, 5.025/$H({\cdot;\,\thetaDT^{(M)}})$} {\draw[color=white, rounded corners=3pt, fill=white, thick] (\x, -0.75) rectangle (\x+0.775, 0.75); \node[rotate=90, align=center] at (\x+0.3875, 0) {\footnotesize \eq};}

    \node[rotate=90, align=center] at (1.2+1, 0.35) {\footnotesize $\hat{\vect{s}}^{(1)}$};
    \node[rotate=90, align=center] at (3.75+1, 0.5) {\footnotesize $\hat{\vect{s}}^{(M-1)}$};

    \node at (3.5,0) {\normalsize $\cdots$};

    \filldraw[gray] (1.2+0.775+0.2,0) circle [radius=2pt];
    \filldraw[gray] (3.75+0.775+0.2,0) circle [radius=2pt];

    \node at (0.2, 0) {\footnotesize $\vect{x}$};

    \draw[fill=white, thick, rounded corners=3pt](6.9, -0.4) rectangle (8, 0.25); \node at (7.45, -0.075) {\footnotesize $\mathcal{L}_\text{o}(\thetaDT)$};

    \draw[-{Latex[length=1.5mm]}, thick] (7.45, 0.8) -- (7.45, 0.25);
    \node at (7.45, 1) {\footnotesize $\vect{y}$};

    \node at (6.5, 0.3) {\footnotesize $\hat{\vect{y}}$};

    \draw[-{Latex[length=1.5mm]}, dash pattern=on 4pt off 2pt, thick] (6.9,-0.2) -- (6,-0.2);

    \node[align=center] at (6.5, -0.6) {\footnotesize Update \\[-0.4ex] \footnotesize $\thetaDT$};
    
\end{tikzpicture}%
}

\newcommand{\pino}{
\begin{tikzpicture}

    \draw[color=ForestGreen, fill=ForestGreen!10, rounded corners=3pt, thick]
      (2,-0.8) rectangle (4,0.8);
    \node[align=center] at (2.55, 0.55) {\footnotesize PINO};
    \draw[color=white, rounded corners=3pt, fill=white, thick]
      (2.2, -0.6) rectangle (3.8, 0.3);
    \node at (3, -0.15) {$G(\cdot; \thetaNN)$};
    \node at (4.5, -0.45) {$\hat{s}(z,t)$};

    \draw[-{Latex[length=1.5mm]}, thick] (1.5,-0.4) -- (2.2,-0.4); \node at (1.2, -0.4) {\footnotesize $(z,t)$};
    \draw[-{Latex[length=1.5mm]}, thick] (1.5,0.1)  -- (2.2,0.1);  \node at (1.2,  0.1){\footnotesize $\vect{x}$};

    \filldraw[black] (1.75, 0.1) circle [radius=2pt];

    \draw[color=RedOrange, fill=RedOrange!15, thick, rounded corners=3pt]
      (5,-1) rectangle (8.2,1);

    \foreach \y/\eq in {0.6/${\loss{L}_\text{ic}(\thetaNN)}$,
                        0.0/${\loss{L}_\text{p}(\thetaNN,\thetaphy)}$,
                       -0.6/${\loss{L}_\text{o}(\thetaNN)}$} {
      \draw[color=black, rounded corners=3pt, fill=white, thick]
        (5.2, \y-0.2) rectangle (8.0, \y+0.2);
      \node[align=center] at (6.5, \y) {\footnotesize \eq};
    }

    \draw[-{Latex[length=1.5mm]}, thick] (4.01,-0.15) -- (5,-0.15);

    \draw[-{Latex[length=1.5mm]},thick] (8.5,-0.6) -- (8,-0.6);
    \node at (8.9,-0.6) {\footnotesize $\vect{y}$};

    \draw[-{Latex[length=1.5mm]}, dash pattern=on 4pt off 2pt, thick]
      (5,0.1) -- (4.01,0.1);
    \node[align=center] at (4.5,0.5) {\footnotesize Update \\[-0.4ex] \footnotesize $\thetaNN$};

    \draw[-{Latex[length=1.5mm]}, dash pattern=on 4pt off 2pt, thick]
      (8.0, 0) -- (8.5, 0);
    \node[align=center] at (8.9, 0) {\footnotesize Update \\[-0.2ex] \footnotesize $\thetaphy$};

    \draw[-{Latex[length=1.5mm]}, thick] (1.75,0.1) -- (1.75, 1.2) -- (6.5, 1.2) -- (6.5, 0.8);
\end{tikzpicture}%
}

\newcommand{\timeline}[3][0.15]{%
  \filldraw[gray] (#2, -1.15) circle [radius=1.5pt];
  \node at (#2,-1.55) {\large $\vdots$};
  \filldraw[gray] (#2, -2) circle [radius=1.5pt];
  \node at (#2, -2.25) {\large $\vdots$};
  \filldraw[gray] (#2, -2.75) circle [radius=1.5pt];
  \node at (#2+#1 - 0.1, -3.03) {\footnotesize #3};
}

\newcommand{\timelineLeft}[3][0.15]{%
  \filldraw[gray] (#2, -1.15) circle [radius=1.5pt];
  \node at (#2,-1.55) {\large $\vdots$};
  \filldraw[gray] (#2, -2) circle [radius=1.5pt];
  \node at (#2, -2.25) {\large $\vdots$};
  \filldraw[gray] (#2, -2.75) circle [radius=1.5pt];
  \node[anchor=west] at (#2+#1-0.35, -3.03) {\footnotesize #3};
}

\newcommand{\timelineRight}[3][0.15]{%
  \filldraw[gray] (#2, -1.15) circle [radius=1.5pt];
  \node at (#2,-1.55) {\large $\vdots$};
  \filldraw[gray] (#2, -2) circle [radius=1.5pt];
  \node at (#2, -2.25) {\large $\vdots$};
  \filldraw[gray] (#2, -2.75) circle [radius=1.5pt];
  \node[anchor=east] at (#2-#1+0.45, -3.03) {\footnotesize #3};
}

\newcommand{\pidt}{
\begin{tikzpicture}
    \node at (4.5, 0.15) {$\hat{s}(z,t)$};
    \draw[color=ForestGreen, fill=ForestGreen!10, rounded corners=3pt, thick] (0.8,-3.3) rectangle (6.2,1.5); \node[align=center] at (1.35, 1.25) {\footnotesize PIDT};
    \draw[thick] (0.5, 0) -- (1, 0); \node at (0.2, -2) {$(z,t)$};
    \draw[-{Latex[length=1.5mm]}, thick] (0.5, -2) -- (1, -2); \node at (0.3, 0.0) {\footnotesize $\vect{x}$};

    \draw[thick] (0.5,0) -- (3,0);

    \draw[color=RoyalBlue, fill=RoyalBlue!10, rounded corners=3pt, thick] (1,-0.95) rectangle (6,1);
    \draw[-{Latex[length=1.5mm]}, thick] (1.01,0) -- (1.2,0); \draw[-{Latex[length=1.5mm]}, thick] (3.75+0.775,0) -- (5.025,0); \draw[-{Latex[length=1.5mm]}, thick] (1.2+0.775 ,0) -- (2.475,0);

    \foreach \x/\eq in {1.2/ $H({\cdot;\,\thetaDT^{(1)}})$, 2.475/$H({\cdot;\,\thetaDT^{(2)}})$, 3.75/$H({\cdot;\,\thetaDT^{(M-1)}})$, 5.025/$H({\cdot;\,\thetaDT^{(M)}})$} {\draw[color=white, rounded corners=3pt, fill=white, thick] (\x, -0.75) rectangle (\x+0.775, 0.75); \node[rotate=90, align=center] at (\x+0.3875, 0) {\footnotesize \eq};}

    \node[rotate=90, align=center] at (1.2+1.04, 0.35) {\footnotesize $\hat{\vect{s}}^{(1)}$};
    \node[rotate=90, align=center] at (3.75+1.04, 0.5) {\footnotesize $\hat{\vect{s}}^{(M-1)}$};

    \node at (3.5,0) {\normalsize $\cdots$};

    \draw[-{Latex[length=1.5mm]}, thick] (6.7, -2) -- (6.7, -4.5) -- (6.2, -4.5);
    \node at (6.73, -1.75) {\footnotesize $\hat{s}(z,t)$};

    \filldraw[gray] (1.2+0.775+0.25,0) circle [radius=2pt];
    \filldraw[gray] (3.75+0.775+0.25,0) circle [radius=2pt];

    \draw[color=Goldenrod, fill=Goldenrod!10, rounded corners=3pt, thick] (1,-2.75) rectangle (6,-1.15);

    \draw[Goldenrod, line width=1pt] (2.225, -1.15) -- (1.2+0.775+0.25,-2.75);
    \draw[Goldenrod, line width=1pt] (4.775, -1.15) -- (3.75+0.775+0.25,-2.75);
    \draw[Goldenrod, line width=1pt] (3.5, -1.15) -- (3.5,-2.75);
    \draw[Goldenrod, line width=1pt] (1, -2) -- (6, -2);

    \draw[-{Latex[length=1.3mm]}, black, line width=0.5pt] (1, -2.75) -- (6.2, -2.75);
    \draw[-{Latex[length=1.3mm]}, black, line width=0.5pt] (1, -2.75) -- (1, -0.92);
    \node[color=black] at (0.9, -1.06) {$t$}; \node[color=black] at (6.1, -2.6) {$z$};

    \timeline{2.225}{${\hat{\vect{s}}^{(1)}}$}
    \timeline{3.5}{${\hat{\vect{s}}^{(m)}}$}
    \timeline{4.775}{${\hat{\vect{s}}^{(M-1)}}$}
    \timelineLeft{1}{$\vect{s}^{(0)}$}
    \timelineRight{6}{$\hat{\vect{s}}^{(M)}$}

    \node at (2.863, -2) {\large $\ldots$};
    \node at (4.113, -2) {\large $\ldots$};

    \draw [BrickRed, thick]  (5.5, -1.75) circle(4 pt);
    \filldraw[BrickRed] (5.5, -1.75) circle [radius=2pt];

    \draw[black, -{Latex[length=1.5mm]}, thick]
  (5.65, -1.75) -- (6.4, -1.75); \node[color=black, fill=white, fill opacity=0.7, text opacity=1] at (5.45, -1.4) {\footnotesize Interpolation}; 
  \node at (1.25, -1.4) {\footnotesize $\hat{\vect{S}}$};
    
    \draw[gray, -{Latex[length=1.5mm]}, line width=1.2pt] (1.2+0.775+0.25, 0) -- (1.2+0.775+0.25,-1.15); \draw[gray, -{Latex[length=1.5mm]}, line width=1.2pt] (3.75+0.775+0.25,0) -- (3.75+0.775+0.25,-1.15);    

    \draw[color=RedOrange, fill=RedOrange!15, rounded corners=3pt, thick] (0.8,-5.2) rectangle (6.2,-3.8);

    \foreach \y/\eq in {-4.4/${\loss{L}_\text{o}(\thetaDT)}$,
                        -5/${\loss{L}_\text{p}(\thetaDT,\thetaphy)}$} {
      \draw[color=black, rounded corners=3pt, fill=white, thick]
        (1, \y) rectangle (6, \y+0.4);
      \node[align=center] at (3.5, \y+0.2) {\footnotesize \eq};
    }

    \draw[-{Latex[length=1.5mm]}, dash pattern=on 4pt off 2pt, thick]
      (1.0, -4.8) -- (0.5, -4.8); \node[align=center] at (0.15, -4.8) {\footnotesize Update \\[-0.2ex] \footnotesize $\thetaphy$};

    \draw[-{Latex[length=1.5mm]}, thick] (0.5, -4.2) -- (1, -4.2);
    \node at (0.25, -4.2) {$\vect{y}$};

    \draw[-{Latex[length=1.5mm]}, dash pattern=on 4pt off 1pt, thick]
      (3.5, -3.8) -- (3.5, -3.3);
    \node at (4.45, -3.55) {\footnotesize Update $\thetaDT$};
\end{tikzpicture}%
}

\newcommand{\betagamma}[1][0.11]{
  \begingroup
  \centering 
  \tikzset{every node/.append style={font=\footnotesize}}
  \pgfplotsset{table/search path={figures/fig_data}}
  \pgfplotsset{
    compat=1.18,
    every axis/.append style={
      scale only axis=false,      
      grid=both,
      minor grid style={opacity=0.15},
      major grid style={densely dotted},
      tick align=inside,
      tick style={semithick},
      xlabel style={yshift=5pt, font=\footnotesize},
      ylabel style={yshift=-6pt, font=\footnotesize},
      scaled x ticks=false,
      scaled y ticks=false,
      xticklabel style={/pgf/number format/fixed,/pgf/number format/precision=1, font=\footnotesize},
      yticklabel style={/pgf/number format/fixed,/pgf/number format/precision=2, font=\footnotesize},
      label style={font=\footnotesize},
      tick label style={font=\footnotesize},
      legend style={font=\footnotesize},
      legend cell align=left, 
      legend columns=3,
      /tikz/column sep=2pt,
      /tikz/row sep=0pt,
    },
    every axis plot/.append style={
      line width=0.65pt,
      mark size=2.2pt,
      mark repeat=15
      mark options={solid, fill=white},
    },
    legend style={font=\footnotesize}
  }

  \pgfplotscreateplotcyclelist{pbmstyles}{%
    {blue!80!black,   solid,           mark=o},
    {red!80!black,    solid,           mark=o},
    {green!60!black,    solid,           mark=o},
    {blue!80!black,   densely dashed,  mark=x, mark options={solid}}, 
    {blue!90!black,    densely dashed,  mark=asterisk, mark options={solid}}, 
    {red!80!black,    densely dashed,  mark=asterisk, mark options={solid}},
  }%
  \pgfplotsset{cycle list name=pbmstyles}

  \pgfmathsetlengthmacro{\GroupW}{\linewidth}
  \pgfmathsetlengthmacro{\hsep}{#1*\GroupW}
  \pgfmathsetlengthmacro{\AxisW}{(\GroupW)/2}
  \pgfmathsetlengthmacro{\AxisH}{0.82*\AxisW}

  \begin{tikzpicture}
    \begin{groupplot}[
      group style={group size=2 by 1, horizontal sep=\hsep},
      width=\AxisW, height=\AxisH,
      legend style={font=\footnotesize},
      legend cell align=left, 
    ]

      \nextgroupplot[
        xlabel={Iteration $\times 10^5$},
        ylabel={Estimated $\hat{\gamma} \times 10^{-3}\,(\mathrm{W\cdot m})^{-1}$},
        legend style={at={(1.15,1.0)}, anchor=south, draw=none, fill=white, fill opacity=0.5, text opacity=1},
        ymin=0.4, ymax=1.48, xmin=0, xmax=3,
        after end axis/.code={%
        },
      ]
        \addplot table [x expr=\thisrow{x}/1e5, y expr=\thisrow{y}/1e-3, col sep=comma] {gamma_pidt_32.csv}; \addlegendentry{PIDT-32};
        \addplot table [x expr=\thisrow{x}/1e5, y expr=\thisrow{y}/1e-3, col sep=comma] {gamma_pidt_64.csv}; \addlegendentry{PIDT-64};
        \addplot table [x expr=\thisrow{x}/1e5, y expr=\thisrow{y}/1e-3, col sep=comma] {gamma_pidt_512.csv}; \addlegendentry{PIDT-512};
        \addplot table [x expr=\thisrow{x}/1e5, y expr=\thisrow{y}/1e-3, col sep=comma] {gamma_pino4_32.csv}; \addlegendentry{$\pinosmall$-32};
        \addplot table [x expr=\thisrow{x}/1e5, y expr=\thisrow{y}/1e-3, col sep=comma] {gamma_pino2_32.csv}; \addlegendentry{$\pinolarge$-32};
        \addplot table [x expr=\thisrow{x}/1e5, y expr=\thisrow{y}/1e-3, col sep=comma] {gamma_pino4_64.csv}; \addlegendentry{$\pinolarge$-64};
        
        \addplot [dashed, black, very thin, forget plot]
          table [x expr=\thisrow{x}/1e5, y expr=\thisrow{y}/1e-3, col sep=comma] {gamma_truth.csv};

        \node[anchor=north west, font=\footnotesize,
              fill=white, fill opacity=0.85, text opacity=1,
              inner sep=1.2pt, rounded corners=1pt]
             at (rel axis cs:0.03,0.99)
             {$\gamma = 1.315\times10^{-3}$ };

      \nextgroupplot[
        xlabel={Iteration $\times 10^{5}$},
        ylabel={Estimated $\hat{\beta}_{2} \times 10^{-26}\,\mathrm{s^2/m}$},
        ymax=-1.65, ymin=-2.12, xmin=0, xmax=3,
        after end axis/.code={%
        },
      ]
        \addplot table [x expr=\thisrow{x}/1e5, y expr=\thisrow{y}/1e-26, col sep=comma] {beta2_pidt_32.csv};
        \addplot table [x expr=\thisrow{x}/1e5, y expr=\thisrow{y}/1e-26, col sep=comma] {beta2_pidt_64.csv};
        \addplot table [x expr=\thisrow{x}/1e5, y expr=\thisrow{y}/1e-26, col sep=comma] {beta2_pidt_512.csv};
        \addplot table [x expr=\thisrow{x}/1e5, y expr=\thisrow{y}/1e-26, col sep=comma] {beta2_pino4_32.csv};
        \addplot table [x expr=\thisrow{x}/1e5, y expr=\thisrow{y}/1e-26, col sep=comma] {beta2_pino2_32.csv};
        \addplot table [x expr=\thisrow{x}/1e5, y expr=\thisrow{y}/1e-26, col sep=comma] {beta2_pino4_64.csv};
        \addplot [dashed, black, very thin, forget plot]
          table [x expr=\thisrow{x}/1e5, y expr=\thisrow{y}/1e-26, col sep=comma] {beta2_truth.csv};

        \node[anchor=north west, font=\footnotesize,
              fill=white, fill opacity=0.85, text opacity=1,
              inner sep=1.2pt, rounded corners=1pt]
             at (rel axis cs:0.05, 0.15)
             {$\beta_{2} = -2.047\times10^{-26}$};

    \end{groupplot}
  \end{tikzpicture}
  \endgroup
}

\newcommand{\profile}[1][]{%
  \begingroup
  \centering 
  \tikzset{every node/.append style={font=\footnotesize}}
  \pgfplotsset{table/search path={figures/profile_data}}
  \pgfplotsset{
    compat=1.18,
    set layers=axis on top,
    every axis/.append style={
      scale only axis=false,    
      grid=both,
      minor grid style={opacity=0.15},
      major grid style={densely dotted},
      tick align=inside,
      tick style={semithick},
      xlabel style={yshift=5pt, font=\footnotesize},
      ylabel style={yshift=-6pt, xshift=-3pt, font=\footnotesize},
      tick label style={font=\footnotesize},
      label style={font=\footnotesize},
      scaled x ticks=false,
      scaled y ticks=false,
      clip marker paths=true,
      legend style={font=\footnotesize},
      legend cell align=left, 
    },
    every axis plot/.append style={
      line width=0.65pt,
      mark size=2.2pt,
      mark options={solid, fill=white},
    },
  }

  \colorlet{pidtcol}{blue!70!black}
  \colorlet{noplcol}{blue!80!black}
  \colorlet{truthcol}{black}

  \pgfmathsetlengthmacro{\AxisW}{\linewidth}
  \pgfmathsetlengthmacro{\AxisH}{\AxisW}

  \begin{tikzpicture}
    \begin{axis}[
      width=\AxisW, height=\AxisH,
      xmin=20, xmax=80,
      xtick={20,40,60,80},
      xlabel={$z$ [km]},
      ylabel={Estimated $\{{\beta}_2^{(m)}\}_{m=1}^M \times 10^{-26} \,\mathrm{s^2/m}$},
      legend style={
        at={(0.5,0.98)}, anchor=south,
        draw=none, fill=white, fill opacity=0, text opacity=1,
        font=\footnotesize, inner sep=2pt,
        /tikz/column sep=5pt, /tikz/row sep=-2pt,
      },
      legend columns=1,
      legend cell align=left, 
      /tikz/column sep=2pt,
      /tikz/row sep=0pt,
    ]

      \node[anchor=south west, font=\footnotesize,
            fill=white, fill opacity=0.8, text opacity=1,
            inner sep=1pt, rounded corners=1pt]
        at (axis cs:21,-1.8e-26/1e-26) {$\beta_2 = -2.047\times10^{-26}$};

      \addplot+[pidtcol, solid, mark=o]
        table [x index=0, y expr=\thisrowno{1}/1e-26, col sep=comma] {beta_PIDT_M4.csv};
      \addlegendentry{PIDT-32}

      \addplot+[noplcol, densely dashed, mark=x]
        table [x index=0, y expr=\thisrowno{1}/1e-26, col sep=comma] {beta_PIDT-no-Lp_M4.csv};
      \addlegendentry{PIDT-32 w/o $\mathcal{L}_p$}

      \addplot+[truthcol, dashed, mark=none, line width=1.2pt]
        table [x index=0, y expr=\thisrowno{1}/1e-26, col sep=comma] {beta_truth.csv};

    \end{axis}
  \end{tikzpicture}
  \endgroup
}

\usepackage{opticameet3} 
\newcommand\authormark[1]{\textsuperscript{#1}}
\usepackage{amsmath,amssymb}
\usepackage[colorlinks=true,bookmarks=false,citecolor=blue,urlcolor=blue]{hyperref} 

\newcommand{\vect}[1]{\mathbf{#1}}
\newcommand{\set}[1]{\mathcal{#1}}
\newcommand{\mat}[1]{\mathbf{#1}}
\newcommand{\loss}[1]{\mathcal{#1}}

\newcommand{\Ltot}{\mathcal{L}}
\newcommand{\Lobs}{\mathcal{L}_{\text{o}}}
\newcommand{\Lic}{\mathcal{L}_{\text{ic}}}
\newcommand{\Lpinn}{\mathcal{L}_{\text{p}}}
\newcommand{\pinosmall}{\text{PINO}_{\text{small}}}
\newcommand{\pinolarge}{\text{PINO}_{\text{large}}}
\newcommand{\thetaDT}{\theta_\text{DT}}
\newcommand{\thetaNN}{\theta_\text{NN}}
\newcommand{\thetaphy}{\hat{\theta}}

\begin{document}

\title{PIDT: Physics-Informed Digital Twin for Optical Fiber Parameter Estimation}

\vspace{-0.4cm}
\author{Zicong Jiang,\authormark{1,*} Magnus Karlsson,\authormark{2} Erik Agrell,\authormark{1} and Christian Häger\authormark{1}}

\address{\authormark{1}Dept. of Electrical Engineering, Chalmers Univ. of Technology, Sweden\\
\authormark{2}Dept. of Microtechnology and Nanoscience, Chalmers Univ. of Technology, Sweden}

\email{\authormark{*}zicongj@chalmers.se} 



\vspace{-0.5cm}
\begin{abstract}
We propose physics-informed digital twin (PIDT): a fiber parameter estimation approach that combines a parameterized split-step method with a physics-informed loss. PIDT improves accuracy and convergence speed with lower complexity compared to previous neural operators. \textcopyright2025 The Author(s)
\end{abstract}

\vspace{0.3cm}
\section{Introduction}
In optical fiber communication systems, accurate knowledge of fiber parameters such as attenuation, dispersion and nonlinearity, and their variation along the link is essential for tasks such as impairment compensation, performance monitoring, and system optimization. 
Recently, learning-based parameter estimation methods have been proposed that embed physics knowledge into training. 
For example, physics-informed neural networks (PINNs) \cite{raissiPhysicsinformedNeuralNetworks2019} incorporate partial differential equation constraints, such as the nonlinear Schrödinger equation (NLSE), as a physics-informed loss. 
PINNs have been applied to optical channel modeling \cite{zangPrincipleDrivenFiberTransmission2022} and estimating fiber parameters \cite{jiangPhysicsInformedNeuralNetwork2022}, but they require retraining for each new initial condition (i.e., transmitted signal). To overcome this problem, neural operator methods have been proposed \cite{lu2021learning}, which learn a solution operator over a distribution of initial conditions for better generalization. 
Variants of these methods, called physics-informed neural operator (PINO) \cite{li2024physics}, have been applied to fiber parameter estimation \cite{songPhysicsInformedNeuralOperatorbased2023}, and channel modeling \cite{songPhysicsInformedNeuralOperator2022, luo2024fast}.
However, prior works mainly use simplified or short signals to simplify training. Moreover, both PINN and PINO rely on conventional neural network (NN) models that require extensive hyperparameter tuning (e.g., number of layers or neurons) and many training iterations to converge.


We propose physics-informed digital twin (PIDT): a neural operator method for accurate and fast parameter estimation with realistic signal sequences. 
PIDT replaces the conventional NN with an interpretable physics-based model, effectively combining the parameterized split-step Fourier method (SSFM)\cite{hager2018nonlinear, hagerPhysicsBasedDeepLearning2021} with a physics-informed loss\cite{raissiPhysicsinformedNeuralNetworks2019}.
A key technical challenge is that, unlike a NN, the SSFM produces outputs that are discrete in space and time. 
This is not directly compatible with the neural operator framework. 
To address this, we introduce a differentiable interpolation step that links SSFM outputs to the physics-informed loss. 
We show that PIDT significantly improves estimation accuracy and convergence speed with lower complexity compared to previous PINO approaches in \cite{songPhysicsInformedNeuralOperatorbased2023}. 
Moreover, PIDT also addresses a limitation of prior SSFM-based parameter estimation approaches in, e.g., \cite{sasai2021digital}, which rely solely on minimizing the observation loss at the fiber output. 
By applying the physics-informed loss to the full $t$- and $z$-dependent optical wavefield, PIDT enhances physical consistency of the mathematical twin model at intermediate fiber locations compared to prior SSFM-based estimation methods. 

\section{Background and related approaches}
\begin{figure}[htbp]
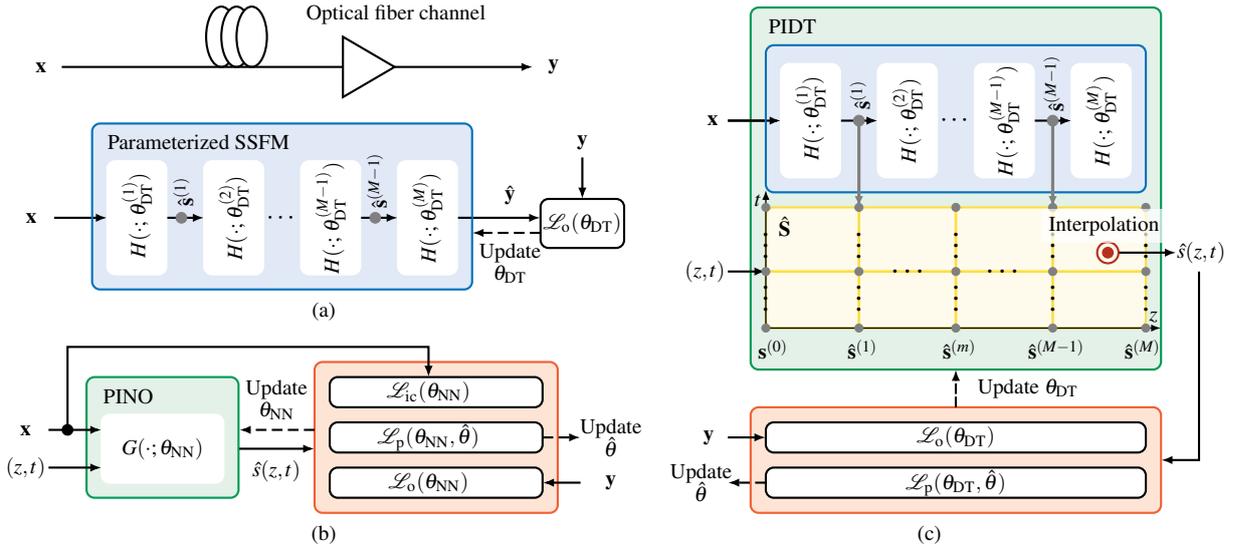

    \footnotesize
  \centering
  \begin{subfigure}[b]{0.55\textwidth}
    \centering
    \begin{subfigure}[b]{\linewidth}
      \centering
      \pbm
      \vspace{-0.1cm}
      \caption{}
      \label{fig:bg1}
    \end{subfigure}
    \begin{subfigure}[b]{\linewidth}
      \centering
      \pino
      \vspace{-0.1cm}
      \caption{}
      \label{fig:bg2}
    \end{subfigure}
  \end{subfigure}
  \hfill
  \begin{subfigure}[b]{0.44\textwidth}
    \centering
    \pidt
    \vspace{-0.1cm}
    \caption{}
    \label{fig:bg3}
  \end{subfigure}
  \vspace{-0.7cm}
  \caption{Fiber-parameter estimation using (a) the parameterized SSFM, (b) PINO, and (c) PIDT (proposed).}
  \vspace{-0.7cm}
  \label{fig:bg}
\end{figure}
Following the convention in the PINO literature, we define the NLSE in residual form $R\left(s(z,t); \theta\right) = 0$, where
\setlength{\abovedisplayskip}{3pt}
\setlength{\belowdisplayskip}{3pt}
\begin{equation}\label{eq:nlse} 
R\left(s(z,t); \theta\right) = \frac{\partial s(z,t)}{\partial z}
+\frac{\alpha}{2} s(z,t)
+\frac{\mathrm{j}\beta_2}{2} \frac{\partial^2 s(z,t)}{\partial t^2}
-\mathrm{j}\gamma |s(z,t)|^2 s(z,t),
\end{equation}
$s(z,t) \in \mathbb{C}$ is the optical baseband signal at space–time coordinate $(z,t)\in [0,L ] \times \mathbb{R}$, $L$ is the propagation distance, and $\theta = \{\alpha$, $\beta_2$, $\gamma \}$ are the attenuation, dispersion, and nonlinear parameters.
The parameter estimation problem is to estimate the coefficients in $\theta$ (or their profiles along the fiber) from observed input--output signal pairs. 
Since our method builds on two existing approaches, we first review SSFM-based and PINO-based parameter estimation. 

\medskip

\noindent\emph{2.1. SSFM-based parameter estimation:}
The SSFM approximates \eqref{eq:nlse} using $M$ segments of length $\Delta z=L/M$, where $\hat{\vect{s}}^{(m)}= H(\hat{\vect{s}}^{(m-1)};\,\thetaDT^{(m)}) = \mat{D}\sigma(\mat{D} \hat{\vect{s}}^{(m-1)})\in\mathbb{C}^N$ is the approximated signal at $z_m=m\Delta z$ starting with $\hat{\vect{s}}^{(0)} = [s(0,t_1), \ldots, s(0,t_N)]^\top$, $t_n=n\Delta t$, $\Delta t=T/N$ is the sampling period, $T$ the time window, and $N$ the number of samples. Here, $\mat{D}=\mat{F}^{-1} \operatorname{diag}\!\big(e^{(-\alpha^{(m)} / 2+\mathrm{j} \beta_2^{(m)} \omega_k^2 / 2) \Delta z/2}\big) \mat{F}$, $\sigma(x)=x e^{\mathrm{j} \gamma^{(m)} \Delta z|x|^2}$ is applied element-wise, $\thetaDT^{(m)} = \{\alpha^{(m)}, \beta_2^{(m)}, \gamma^{(m)}\}$ are segment-wise tunable parameters, $\mathbf{F}$ is the discrete Fourier transform matrix (periodic boundary conditions are assumed), and $\omega_k$ the $k$-th angular frequency. We denote the overall SSFM mapping as $\hat{\vect{y}}(\hat{\vect{s}}^{(0)}; \thetaDT) = \hat{\vect{s}}^{(M)}$, where $\thetaDT=\{ \thetaDT^{(m)} \}_{m=1}^M$. For parameter estimation (see Fig.~\ref{fig:bg1}), $\thetaDT$ is optimized by minimizing the observation loss $\Lobs(\thetaDT)=(1/|\set{D}|) \sum_{(\vect{x}, \vect{y} ) \in \set{D}}\|\hat{\vect{y}}(\vect{x}; \thetaDT)-\vect{y}\|^2$, where $\mathcal{D}$ is the training set consisting of pairs of input and received reference signals. After training, $\thetaDT$ can be directly extracted from the model. However, since $\Lobs$ constrains only the output, it may yield parameter sets that violate physical consistency along the fiber. This motivates the use of additional loss functions that constrain the full wavefield.

\medskip

\noindent\emph{2.2. PINO-based parameter estimation:}
The PINO in Fig.~\ref{fig:bg2} learns a solution operator $G(\vect{x}, z, t; \thetaNN)$ that maps input signals $\vect{x} \in \mathbb{C}^N$ and space-time coordinates $(z,t)$ to the approximated signal $\hat{s}(z,t) \in \mathbb{C}$. In prior works~\cite{songPhysicsInformedNeuralOperator2022,songPhysicsInformedNeuralOperatorbased2023}, $G$ is implemented as a black-box NN\footnote{The mapping is typically divided into a branch and a trunk network, with complex signals separated into real and imaginary parts.} with parameters $\thetaNN$ (weights and biases), so fiber parameters cannot be directly extracted from it. 
Instead, fiber parameters $\thetaphy=\{\hat{\alpha}, \hat{\beta}_2,\hat{\gamma}\}$ are estimated by minimizing the physics-informed loss $\Lpinn(\thetaNN,\thetaphy)=(1/(|\set{X}||\set{C}|)) \sum_{\vect{x} \in \set{X}} \sum_{(z,t)\in \set{C}}|R(G(\vect{x}, z,t; \thetaNN);\thetaphy)|^2$, where $\mathcal{X} = \{ \vect{x} \,|\, (\vect{x}, \vect{y}) \in \mathcal{D}\}$ and $\mathcal{C}$ is a set of predefined space-time coordinates. 
However, training only with $\mathcal{L}_\text{p}$ may yield incorrect $\thetaphy$, as the operator $G$ can learn to satisfy the NLSE residual $R$ with mismatched parameters. 
To address this, $\Lpinn$ is combined with an initial-condition loss $\Lic(\thetaNN)=(1/|\set{X}|) \sum_{\vect{x} \in \set{X}}\|\hat{\vect{x}}(\vect{x}; \thetaNN)-\vect{x}\|^2$ and an observation loss $\Lobs(\thetaNN) = (1/|\mathcal{D}|) \sum_{(\vect{x}, \vect{y}) \in \mathcal{D} } \| \hat{\vect{y}}(\vect{x}; \thetaNN) - \vect{y} \|^2$, where $\hat{\vect{x}}(\cdot; \thetaNN) = [ G(\cdot, 0, t_1; \thetaNN), \ldots,  G(\cdot, 0, t_N; \thetaNN)]^\top$ and $\hat{\vect{y}}(\cdot; \thetaNN) = [ G(\cdot, L, t_1; \thetaNN), \ldots,  G(\cdot, L, t_N; \thetaNN)]^\top$. 
The total loss is
$\Ltot(\thetaNN,\thetaphy)=\lambda_\text{o}\Lobs(\thetaNN)
+\lambda_\text{ic}\Lic(\thetaNN)
+\lambda_\text{p}\Lpinn(\thetaNN,\thetaphy)$, where
$\lambda_\text{o},\lambda_\text{ic}, \lambda_\text{p}$ are adjustable weights. 

\section{Proposed physics-informed digital twin (PIDT) for fiber parameter estimation}
While neural operator methods like PINO can enable parameter estimation across different input signals, learning the full NLSE solution operator remains challenging: conventional NN models are sensitive to architecture choices and need many training iterations to converge. To overcome these issues, we propose PIDT, which embeds the parameterized SSFM as a differentiable, physics-based twin model within the operator-learning framework. 

The main challenge is that PINO requires predictions at arbitrary coordinates, whereas the SSFM produces outputs on a discrete space--time grid. We resolve this interface mismatch through a differentiable interpolation step, as illustrated in Fig.~\ref{fig:bg3}.
In PIDT, the proposed neural operator has two inputs: the discretized transmitted signal $\vect{x}$ and a querying coordinate $(z,t)$, consistent with the PINO structure in Fig.~\ref{fig:bg2}. Leveraging the SSFM, we obtain the output signal vector $\hat{\vect{s}}^{(m)}$ at the end of each segment $m$ and combine them with the initial condition $\hat{\vect{s}}^{(0)} = \vect{x}$ to form an intermediate signal matrix
$\hat{\mat{S}}=
[\,\hat{\vect{s}}^{(0)},\,\ldots,\,\hat{\vect{s}}^{(m)},\,\ldots,\,\hat{\vect{s}}^{(M)}\,] \in \mathbb{C}^{N\times (M+1)}$. 
The approximated signal value $\hat{s}(z,t)$ at arbitrary coordinates $(z,t)$ is then obtained by appropriately interpolating the matrix $\hat{\mat{S}}$, where the choice of interpolation method is nontrivial in general. For example, a simple piecewise linear interpolation in time yields a zero second-order derivative, preventing gradient flow to the dispersion parameter $\hat{\beta}_2$. In this paper, we employ natural bicubic splines \cite{hall1973natural}, which ensure continuous second-order derivatives across intervals.

Since PIDT follows the neural-operator formalism, the loss functions are identical to those in Sec.~2.2. However, owing to the physics-based model and interpolation, the initial condition is inherently satisfied at all coordinates $(z=0, t_n)$, and no additional $\Lic$ term is required. 
Moreover, after training, physically meaningful and interpretable parameters appear both in the SSFM twin model via $\thetaDT$ (where they govern signal propagation), and in the physics-informed loss via $\thetaphy$ (where they define the NLSE residual for consistency enforcement). 

\begin{figure}[t]
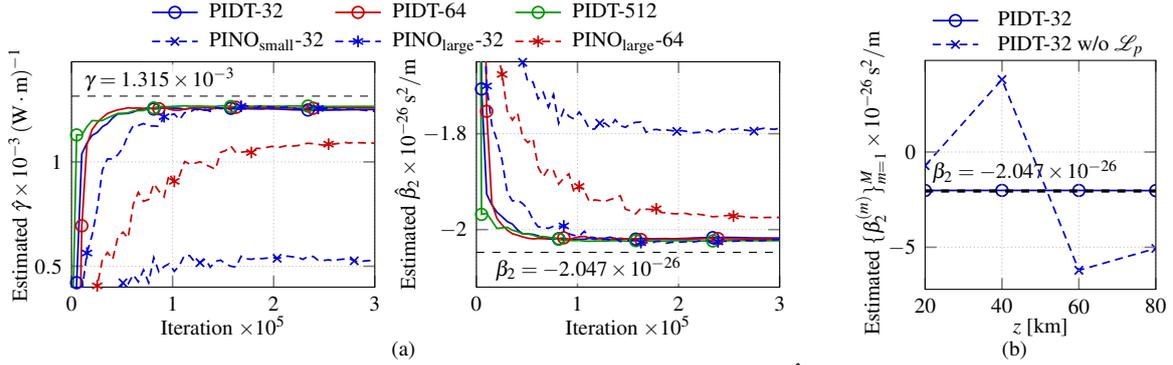

  \centering
  \begin{subfigure}[b]{0.7\linewidth}
    \centering
    \betagamma[0.12]
    \vspace{-0.3cm}
    \subcaption{}
    \label{fig:betagamma}
  \end{subfigure}
  \hfill
  \centering
  \begin{subfigure}[b]{0.29\linewidth}
    \centering
    \profile
    \vspace{-0.3cm}
    \subcaption{}
    \label{fig:profile}
  \end{subfigure}
  \vspace{-1.1cm}
  \caption{Estimation results comparing PIDT-$N_\text{sym}$ $(M=4)$ with (a) PINO-$N_\text{sym}$ for $\thetaphy$ and (b) parameterized SSFM for $\thetaDT$. Both $\pinosmall$ ($3$ layers, $200$ neurons/layer) and $\pinolarge$ ($7$ layers, $900$ neurons/layer) use \texttt{tanh} activation. }
  \vspace{-0.7cm}
  \label{fig:pidt_pino_summary}
\end{figure}

\section{Numerical Results}
To enable comparison with prior work, we adopt a similar setup as in~\cite{songPhysicsInformedNeuralOperatorbased2023}: $N_\text{sym}$ 16-QAM symbols at 14~Gbaud are pulse-shaped (root-raised cosine, roll-off 0.1, oversampling $\rho=2$, launch power $P_0=0\,$dBm) and transmitted over a fiber of length $L=80$~km, resulting in $N=\rho N_\text{sym}$ samples. 
Attenuation is neglected ($\alpha=0$) and only $\beta_2$ and $\gamma$ are estimated, following \cite{songPhysicsInformedNeuralOperatorbased2023}. 
Reference signals are generated using the SSFM with $M=800$, 
where noise with power $P_\text{n} = -20\,$dBm is added at $L=80$\,km in each training iteration, giving $\text{SNR}=10\log_{10}(P_0/P_\text{n})=20\,$dB. For the neural operator in PIDT, we set $M=4$. 
For PINO, several NN architectures for $G$ are evaluated, as discussed in the following. Since no initialization scheme was specified in \cite{songPhysicsInformedNeuralOperatorbased2023}, we initialize the parameters in $\thetaDT,\,\thetaNN$ as $\mathcal{N}(0,1)$ (following \cite{lu2021learning}) and parameters in $\thetaphy$ to 0 (following~\cite{jiangPhysicsInformedNeuralNetwork2022}).  Loss weights $\lambda_\text{ic},\lambda_\text{o},\lambda_\text{p}$ are adaptively updated using the balancing scheme in~\cite{wang2023expert}. Other implementation details are provided in the code~\cite{Jiang2025PIDT}.



Figure~\ref{fig:betagamma} compares parameter estimation accuracy for PINO and the proposed PIDT. For $N_\text{sym}=32$ (the longest symbol sequence considered in \cite{songPhysicsInformedNeuralOperatorbased2023}), PINO performance strongly depends on the NN architecture: a smaller model ($\pinosmall$) performs poorly, while larger configurations improve accuracy but require careful tuning. Even with an optimized architecture ($\pinolarge$), PIDT achieves slightly higher accuracy and converges in fewer iterations.  Increasing $N_\text{sym}$ to $64$, the performance of $\pinolarge$ degrades, and further enlarging the NN in our simulations did not yield any improvement. 
In contrast, PIDT–64 maintains high accuracy without any tuning, and continues to perform reliably even for $N_\text{sym}=512$, highlighting its potential for handling longer signal sequences.


\newcommand{\Cpidt}{C_\text{PIDT}}
\newcommand{\Cpino}{C_\text{PINO}}

In terms of complexity, we compare the number of trainable parameters in $G$ and the approximate number of real multiplications per symbol to evaluate $G$. 
For the considered setting, PIDT has $|\thetaDT| = 2M = 8$ trainable parameters in $G$, whereas $\pinosmall$ and $\pinolarge$ contain, respectively $|\thetaNN| \approx 4\times10^5$ and $|\thetaNN| \approx 2 \times 10^7$ weights alone (excluding biases), making training substantially heavier.
For each input signal, PIDT requires $\Cpidt = \left(M(8N\log_2N + cN) + C_\text{int}\right)/N_\text{sym}$ real multiplications per symbol, where $cN\!\approx\!24N$ covers $8N$ nonlinear and $16N$ linear frequency-domain operations, and $C_\text{int}=52$ accounts for the interpolation (assuming pre-calculated interpolation coefficients). In contrast, the number of multiplications per symbol for NN-based PINO, approximated by $\Cpino = |\thetaNN|/N_\text{sym}$, is roughly $\Cpino/\Cpidt \approx 20$ times higher for $\pinosmall$ and $N_\text{sym} = 32$.




To assess physical consistency, Fig.~\ref{fig:profile} shows the estimated $\{\beta_2^{(m)}\}_{m=1}^M$ extracted from PIDT after training with and without the physics-informed loss $\Lpinn$. 
The latter corresponds to training the parameterized SSFM only on the observation loss $\Lobs$. 
While this achieves a smaller output prediction error after training ($\Lobs \approx 6\times10^{-4}$) compared to PIDT ($\Lobs \approx 1.1\times10^{-3}$), the resulting parameter profiles deviate notably from the ground truth. In contrast, the inclusion of $\Lpinn$ in PIDT yields more accurate profiles, confirming the benefit of enforcing physical consistency. 


\vspace{0.1cm}

\noindent\footnotesize{\textbf{Acknowledgements:} The work of C.~H\"ager and E.~Agrell was supported by the Swedish Research Council under grants no.~2020-04718 and 2021-03709, resp. The simulation resources were provided by the National Academic Infrastructure for Supercomputing in Sweden (NAISS), partially funded by the Swedish Research Council through grant agreement no. 2022-06725.}

\vspace{-0.3cm}
\bibliography{sample}
\label{sample}

\end{document}